\documentstyle[sprocl]{article}

\input{psfig.sty}

\def\NPB{{\em Nucl. Phys.} B}
\def\PLB{{\em Phys. Lett.}  B}
\def\PRL{\em Phys. Rev. Lett.}
\def\PRD{{\em Phys. Rev.} D}
\def\ZPC{{\em Z. Phys.} C}

\def\nn{\noindent}
\def\ie{{\it i.e.}}
\def\eg{{\it e.g.}}

\def\etal{{\it et al.}}
\def\be{\begin{equation}}
\def\ee{\end{equation}}
\def\bea{\begin{eqnarray}}
\def\eea{\end{eqnarray}}
\begin{document}

\rightline{\vbox{\halign{&#\hfil\cr
&SLAC-PUB-7936\cr
&September 1998\cr}}}
\vspace{0.8in}

\title{{RARE K DECAYS AND NEW PHYSICS BEYOND THE STANDARD MODEL}
\footnote{To appear in the {\it Proceedings of the Workshop on $CP$ 
Violation}, University of Adelaide, Adelaide, Australia, 3-8 July, 1998}
}

\author{ {T. G. RIZZO}
\footnote{Work supported by the Department of Energy, 
Contract DE-AC03-76SF00515}
}

\address{Stanford Linear Accelerator Center, 
Stanford University, Stanford,\\ CA 94309, USA
\\E-mail: rizzo@slacvx.slac.stanford.edu}


\maketitle\abstracts{ Rare decays provide a complementary window to direct 
collider searches as probes of physics beyond the Standard Model. We present 
an overview of the New Physics sensitivity provided by existing and future 
measurements of rare leptonic and semileptonic $K$ decays within the context 
of several classes of models.}

\section{Introduction}

Rare $K$ decays have historically provided strict constraints on the 
construction of New Physics models. Amongst these many decays those involving 
leptons are the most theoretically clean. Here we will provide an overview of 
what these modes are telling us about the parameter spaces of several New 
Physics scenarios. We can categorize these rare $K$ decays into several 
distinct classes, which we will discuss in turn, as follows:

\begin{itemize}

\item  Non-Standard Model Modes. These consist of a handful of decays such as 
$K_L \to e\mu$, $K\to \pi e\mu$ and $K\to \pi X$ which just don't fly in 
the Standard Model (SM). The simple observation of {\it any} of these modes 
would signal New Physics. 

\item  Precision Modes. These decay modes occur in the SM but have associated 
with them $T$-odd correlations which possibly lead to $CP$-violating 
observables that can only be probed by high precision experiments. Examples 
are the measurements of the transverse $\mu$ polarization in either 
$K\to \pi \mu \nu$ or $K\to \mu \nu \gamma$. 

\item  Short Distance Dominated Modes. These particularly clean modes 
occur in the SM but rate enhancement or suppression would signal New Physics. 
$K^+\to \pi^+ \nu \nu$ and the $CP$-violating $K_L\to \pi^0 \nu \nu$ channels
provide the best examples and are very well understood in the SM. 

\item  Long-Distance Dominated Modes. These include such decays as 
$K\to \ell^+\ell^-$, $K\to \ell^+\ell^- \gamma$ and $K\to \pi \ell^+\ell^-$ 
which provide fertile testing grounds for chiral perturbation 
theory(ChPT)~{\cite {chiral}} but which have somewhat diminished sensitivity to 
New Physics due to long distance uncertainties. (Perhaps $K\to \mu^+\mu^-$ is 
an exception.) We will have nothing to say about these modes in the discussion 
below. 

\end{itemize}

\section{Non-Standard Model Modes}

\begin{figure}[htbp]
\centerline{
\psfig{figure=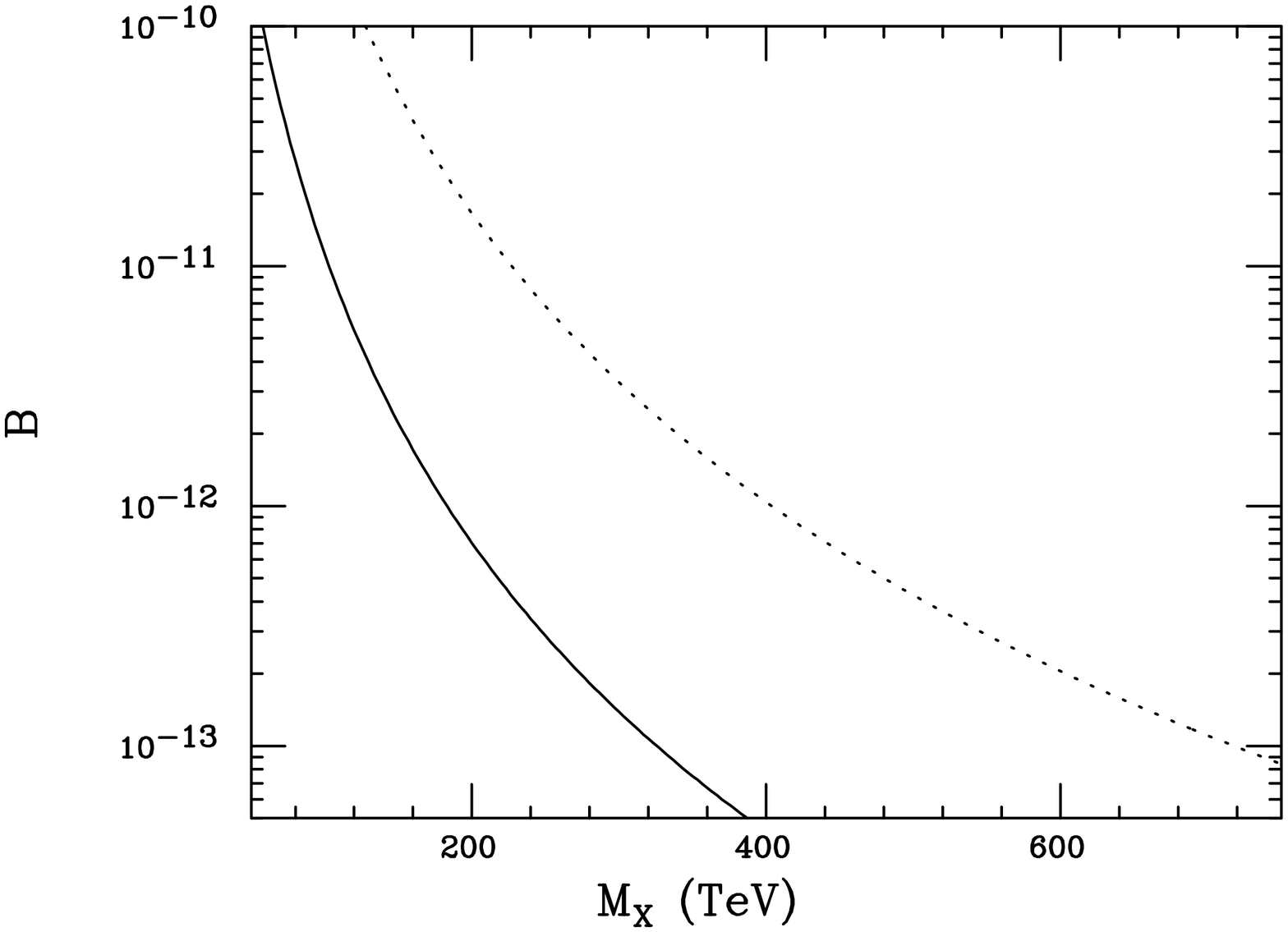,height=8.2cm,width=9.5cm,angle=-90}}
\vspace*{-0.30cm}
\centerline{
\psfig{figure=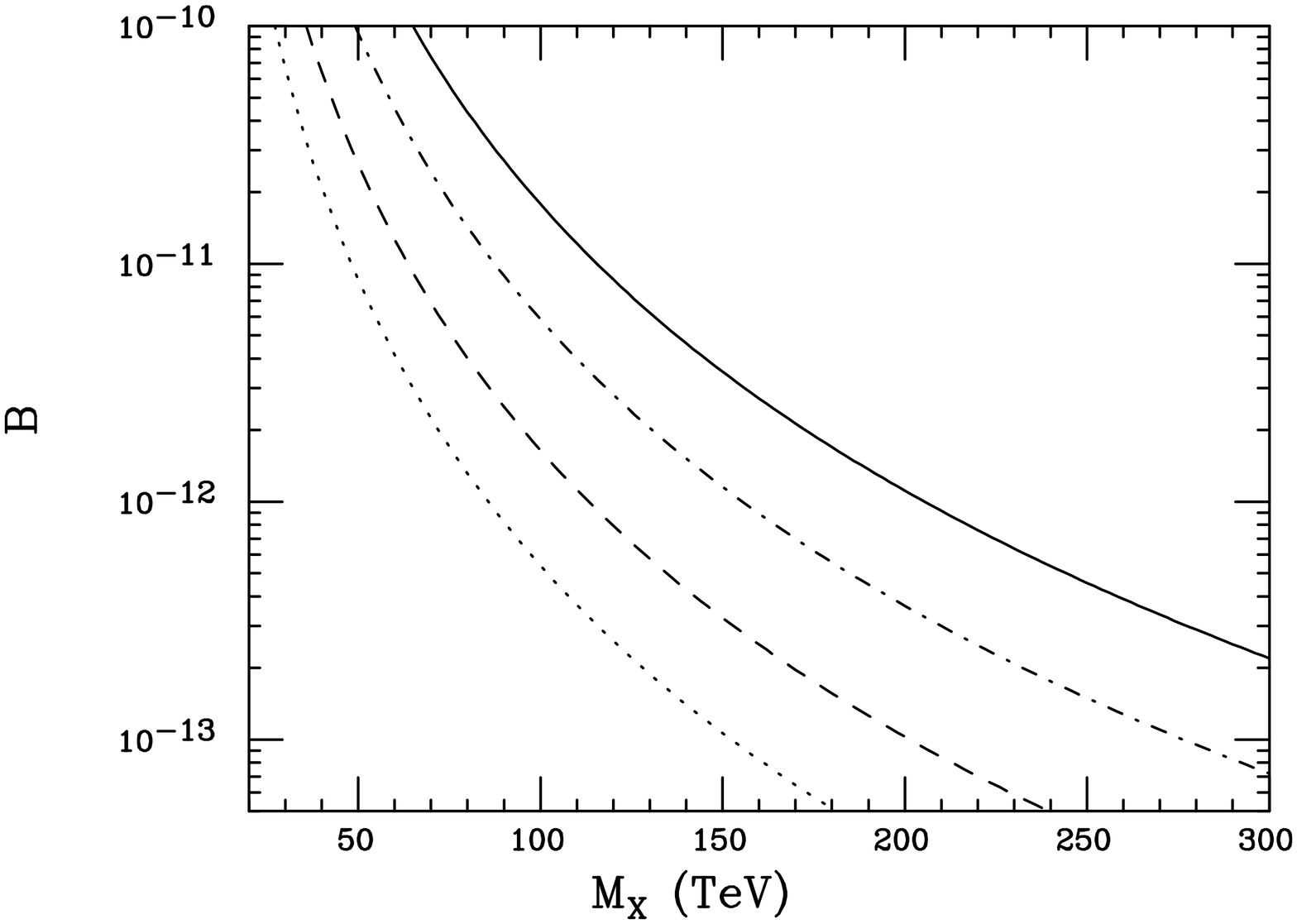,height=8.2cm,width=9.5cm,angle=-90}}
\vspace*{-0.2cm}
\caption{Branching fractions for $K_L \to e\mu$(top) and 
$K\to \pi e\mu$(bottom) as a function of $M_X$ assuming $g_X=g$. In the top 
panel the solid(dotted) curve corresponds to A(P) exchange. 
In the bottom panel, the $K_L$ mode is represented by the solid(S) and 
dashed(V) curves while the $K^+$ mode is represented by the dash-dotted(S) 
and dotted(V) curves.}
\label{fig1}
\end{figure}

Both $K_L \to e\mu$ and $K\to \pi e\mu$ proceed through operators of the form 
\begin{eqnarray}
{\cal O}_{V,A}&=& {g_X^2\over {2M_X^2}}\bar d\gamma_\mu[C_{Lq}P_L+C_{Rq}P_R]s
\cdot \bar \mu\gamma^\mu[C_{L\ell}P_L+C_{R\ell}P_R]e+h.c.\,, \nonumber \\
{\cal O}_{S,P}&=& {g_X^2\over {2M_X^2}}\bar d [C'_{Lq}P_L+C'_{Rq}P_R]s
\cdot \bar \mu[C'_{L\ell}P_L+C'_{R\ell}P_R]e+h.c.\,, 
\end{eqnarray}
but are complementary in that $<0|{\cal O}_{A,P}|K>\neq 0$ and 
$<\pi|{\cal O}_{V,S}|K>\neq 0$ with all other matrix elements being zero. 
As it stands, the form of the above operators conserve generation number. It 
is important to remember that other operators of the same kind may exist 
wherein the roles of $e$ and $\mu$ are interchanged so that generation number 
is no longer conserved. In principle, all operators such as those above 
may be generated either through loops or via 
tree-level exchanges. Branching fractions for the above decays can be 
immediately calculated from these operators and the well-known SM matrix 
elements apart from small isospin corrections, \eg, the branching fraction 
for $K_L \to e\mu$ due to ${\cal O}_A$ is 
\begin{equation}
B_A=11.24\cdot 10^{-12}~\big [{g_X\over {g}} {100 TeV\over {M_W}}\big ]^4 
~(C_{Lq}-C_{Rq})^2~(C_{L\ell}^2+C_{R\ell}^2)\,,
\end{equation}
with qualitatively similar results holding for the other modes and operators, 
apart from chiral enhancement factors in the cases of $S$- and $P$-type 
couplings. 
These results are summarized in Fig.~\ref{fig1} assuming $g_X=g$, the weak 
coupling constant, and the product of the coefficients 
$C_i$'s is set to unity. Since the 
present limit~{\cite {jung}} on the branching fraction for $K_L \to e\mu$ is 
$3\cdot 10^{-12}$ at $90\%$ CL, we see that for tree-level exchanges with 
weak couplings, mass scales $>100$ TeV, beyond the reach of any planned 
collider, are already being probed. Somewhat lower scales, but no less 
impressive, are probed by the corresponding 
$K_L\to \pi^0 e\mu$ and $K^+\to \pi^+ e\mu$ modes with branching fraction 
upper limits~{\cite {jung}} of $3.2\cdot 10^{-9}$ and $4\cdot 10^{-11}$, 
respectively, at the $90\%$ CL. 
Unfortunately, due to the $\sim M_X^4$ scaling of the branching fractions 
for both $K_L \to e\mu$ and $K\to \pi e\mu$ decay modes, the mass range being 
probed will improve quite slowly as bounds on these branching fractions 
improve. 

\vspace*{-0.5cm}
\nn
\vspace*{0.4mm}

$R$-parity violating interactions that also violate lepton number can mediate 
both $K_L \to e\mu$ and $K\to \pi e\mu$ processes through tree-level 
exchanges. The relevant trilinear terms in the superpotential are given by 
$W_R=\lambda_{ijk}L_iL_jE^c_k+\lambda'_{ijk}L_iQ_jD^c_k$
 (where $i,j,k$ are family indices and symmetry demands that $i<j$ in the term
proportional to $\lambda$). In the case of, \eg, $K_L\to e\mu$, the reaction  
occurs through both $s$-channel $\tilde \nu$ exchange(which induces S,P-type 
operators) as well as $t$-channel $\tilde u$ exchange(which induces V,A-type 
operators via a Fierz transformation)~{\cite {candp}}. Interestingly if the 
$\tilde \nu$ 
exchange dominates, both the processes $K_L\to e^+\mu^-$ and 
$K_L\to e^-\mu^+$ can occur but need 
not have equal branching fractions. If the $\tilde \nu (\tilde u)$ channel 
dominates then the very strong existing bound on the branching fraction 
severely constrains the product of the relevant Yukawa couplings to be 
$< 7\cdot 10^{-9}(1\cdot 10^{-7})$ for 100 GeV SUSY particle masses. 

Similarly, leptoquark(LQ) exchange~{\cite {lepto}} can also be responsible for 
these lepton number violating processes by 
inducing V,A-type operators. Using the 
current branching fraction bound results in a roughly model-independent limit 
of $M_{LQ} \geq 150 \sqrt {\lambda \lambda'\over {g^2}}$ TeV where 
$\lambda \lambda'$ is the product of the relevant Yukawas. This is one of the 
major reasons for believing that LQ couplings are family diagonal. 

The two-body decay $K\to \pi X$ where $X$ is invisible can occur in many 
scenarios, \eg, the familon model~{\cite {wilc}}. If one describes such decays 
via an effective interaction of the form 
\begin{equation}
{\cal L}={1\over {F}} \partial^\mu X(\bar u \gamma_\mu s)+h.c.\,,
\end{equation}
then the lack of 
observation of this decay by E-787 {\cite {e787}} at the $90\%$ CL of 
$3.0\cdot 10^{-10}$ implies $F>3.1\cdot 10^{11}$ GeV, an impressively large 
scale.

\section{Precision Modes}

In the absence of final state interactions(FSI) the $T$-odd transverse 
$\mu$ polarization, 
$P_\mu^T \sim \mbox {\bf s}_\mu\cdot (\mbox{\bf p}_{\pi,\gamma}
\times \mbox{\bf p}_\mu)$ in either $K\to \pi \mu \nu$ or 
$K\to \mu \nu \gamma$ is a new measure of $CP$ violation as it is effectively 
zero in the SM. Fortunately, the effects of FSI have been shown to be 
reasonably small (or at least not dangerously large)~{\cite {fsi}} particularly 
for the $K\to \pi$ case. For $K\to \pi \mu \nu$ one expects FSI at the level 
of $\sim 10^{-6}$ whereas for $K\to \mu \nu \gamma$ they are expected to be 
$\sim 10^{-3}$. 
Both processes can be analyzed in a model-independent 
manner using the formalism of Wu and Ng~{\cite {wung}}.

The most general 
dimension-six interaction responsible for $K\to \pi \mu \nu$ assuming only 
left-handed neutrinos can be written as
\begin{equation}
{\cal L}={\cal L}_{SM}+2[\bar u (G_S+G_P\gamma_5) s](\bar \mu P_L\nu)+
2[\bar u \gamma_\mu(G_V+G_A\gamma_5) s](\bar \mu \gamma^\mu P_L\nu)\,,
\end{equation}
but the $A,P$ terms do not contribute to this decay having zero matrix 
elements due to parity conservation. (They will contribute in the case of the 
$K\to \mu \nu \gamma$ decay.) 
From ${\cal L}$ one obtains an effective Hamiltonian 
which has the same structure as in the SM: 
\begin{equation}
{\cal H}_{eff}={G_F\over {\sqrt 2}}V_{us}\bar u(\mu)\gamma^\lambda(1-\gamma_5)
v(\nu)\big [f'_+(p_K+p_\pi)_\lambda+f'_-(p_K-p_\pi)_\lambda\big]\,,
\end{equation}
where $f'_\pm=f_\pm (1+\delta_\pm)$, $f_\pm$ being the well-known SM 
form-factors and 
\begin{eqnarray}
\delta_+ &=& {\sqrt 2 G_V\over {G_F V_{us}}}\,, \nonumber \\
\delta_- &=& \delta_+ +{\sqrt 2 G_S m_K^2\over {G_F V_{us}(m_s-m_d)m_\mu}}
\big[{f_+\over {f_-}}(1-r)+{q^2\over {m_K^2}}\big]\,,
\end{eqnarray}
with $r=(m_\pi/m_K)^2$. A short analysis then shows that for this mode 
$P_\mu^T(\pi) \sim Im (\xi)=Im [f'_-/f'_+] \sim Im G_S$, so that only 
this quantity needs to be calculated in any given model~{\cite {models}}. 
If $G_S$ cannot be generated or if it is real then $P_\mu^T(\pi)$ remains zero. 
The current experimental results summarized by the PDG~{\cite {pdg}} are 
$P_\mu^T(K^+)=-0.0031\pm 0.0053$ and $P_\mu^T(K^0)=0.0021\pm 0.0048$. In 
addition, 
KEK-246 has recently reported~{\cite {kek246}} a new preliminary result of 
$P_\mu^T(K^+)=-0.0025\pm 0.0058$. In the future, 
KEK-246 expects to reach an experimental 
sensitivity of 0.0013 whereas AGS-936 and AGS-923 expect to reach sensitivities 
of 0.00035 and 0.00013, respectively~{\cite {litt}}.

$K\to \mu \nu \gamma$ can be analyzed in a similar manner. The matrix element 
has two contributions, one due to Inner Bremsstrahlung(IB) where the photon 
comes off one of the charged legs or the $K \mu \nu$ vertex, and a Structure 
Dependent(SD) term arising from loops. THese can be written symbolically as 
\begin{eqnarray}
M_{IB} &\sim & f'_K K^\mu \epsilon^*_\mu\,, \nonumber \\
M_{SD} &\sim & L_\nu H^{\mu\nu}\epsilon^*_\mu\,,
\end{eqnarray}
with $L_\nu$ and $K^\mu$ being vectors formed from the leptonic momenta, 
$\epsilon_\mu$ is the photon polarization vector and, 
with $p(q)$ being the $K(\gamma)$ momenta, the hadronic tensor is
\begin{equation}
H^{\mu\nu}=F'_A\big [-g^{\mu\nu}p\cdot q+p^\mu q^\nu \big ]+iF'_V 
\epsilon^{\mu\nu\tau\sigma}q_\tau p_\sigma\,. 
\end{equation}
The New Physics is encoded in the primed quantities: 
$f'_K=f_K(1-\Delta_P-\Delta_A)$, $F'_A=F_A(1-\Delta_A)$ and 
$F'_V=F_V(1+\Delta_V)$, with the unprimed quantities corresponding to 
their SM values. ($F_V=-0.0945$ and $F_A=-0.0425$ in chiral perturbation theory 
at the one-loop level~{\cite {chiral}}.) The $\Delta_i$ are given by 
\begin{equation}
\Delta_{P,A,V}={\sqrt 2\over {G_F V_{us}}}\bigg[{G_P m_K^2\over {
(m_s+m_d)m_\mu}},~G_A,~G_V\bigg]\,,
\end{equation}
with the $G_i$ as defined in the effective Lagrangian above. A short 
analysis then shows that 
$P_\mu^T(\gamma)={\cal F}_1 Im(\Delta_A+\Delta_V)+{\cal F}_2 Im(\Delta_P)$ 
where ${\cal F}_{1,2}$ are phase space factors of order $\sim -0.1$. Here we 
note that $G_S$ does {\it not} contribute; nor will a common overall shift in 
the V-A coupling since then the sum $\Delta_R=\Delta_V+\Delta_A$ will be zero. 
Thus, for example, in the Left-Right Symmetric Model~{\cite {right}}, 
$P_\mu^T(\pi)=0$ but 
$P_\mu^T(\gamma)\sim -2\cdot 10^{-3}~[\phi/10^{-2}]~Im[e^{i\omega}V^R_{us}
/|V^L_{us}|]$, where $\phi(\omega)$ is the $W_L-W_R$ mixing angle(phase) and 
$V^{L,R}$ are the left- and right-handed CKM matrices which need not be 
directly related. As shown by Wu and 
Ng~{\cite {wung,models}}, a significant $\Delta_R$ can also be generated by 
large stop mixing. In models with leptoquarks, $R$-parity violation or an 
enriched Higgs sector~{\cite {wung,models,moremods}}, one 
generates both $Im(G_{S,P})\neq 0$ with 
comparable magnitudes leading to values of $P_\mu^T(\pi, \gamma)$'s in the 
$few \cdot 10^{-3}$ range and with the same sign. This level of polarization 
should be observable in the future. 

This discussion demonstrates that the two modes $K\to \pi \mu \nu$ and 
$K\to \mu \nu \gamma$ 
are complementary in their sensitivity to operators with different tensor 
structures generated by New Physics.

\section{Short-Distance Dominated Modes} 

$K^+\to \pi^+\nu \bar \nu$ and $K_L\to \pi^0 \nu \bar \nu$
are two of the most well understood rare decays in the 
SM~{\cite {bigb}} with anticipated branching fractions of 
$(9.1\pm 3.2)\cdot 10^{-11}$ and $(2.8\pm 1.7)\cdot 10^{-11}$, respectively. 
(Note most of the uncertainty in these values arises from our poor 
knowledge of the input 
parameters and not from theoretical uncertainties.) 
E-787~{\cite {e787}} has recently obtained evidence for the charged decay 
mode with a branching fraction of 
$B=(4.2^{+9.7}_{-3.5})\cdot 10^{-10}$, consistent with the SM but leaving room 
for New Physics. In the case of the neutral mode there is presently only an  
upper bound~{\cite {ktev}} from KTEV which is $B< 1.6\cdot 10^{-6}$ at 
$90\%$ CL. 

In almost all models with only left-handed $\nu$'s the effective Hamiltonian 
describing this decay can be written as 
\begin{equation}
{\cal H}_{eff}= {2G_F\over \sqrt {2}}{\alpha\over {\pi \sin^2 \theta_W}} 
\lambda_t \bar s \gamma^\mu(X_LP_L+X_RP_R)d~\bar \nu_\ell \gamma_\mu 
P_L \nu_\ell~+~h.c.\,,
\end{equation}
with $\lambda_i=V^*_{is}V_{id}$ and $X_{L,R}$ being model-dependent. (Note 
that it is possible that more general coupling structures which include 
scalar and pseudoscalar terms may be generated 
via box diagrams in some more exotic  models not discussed below.) In the 
SM, $X_R=0$ and $X_L=X_t+{\lambda^4 \lambda_c\over {\lambda_t}}P_c$, with the 
top contribution being $X_t\sim 1.5$, $\lambda$ being the usual Wolfenstein 
parameter and $P_c=0.40\pm 0.06$ being the highly suppressed 
charm contribution. From this Hamiltonian one obtains the following branching 
fractions:
\begin{eqnarray}
B(K^+\to \pi^+\nu \bar \nu) &=& 4.11\cdot 10^{-11}
\big|{\lambda_t(X_L+X_R)\over {\lambda^4}}\big|^2\,, \nonumber \\
B(K_L\to \pi^0 \nu \bar \nu) &=& 1.80\cdot 10^{-10}
\big[Im {\lambda_t(X_L+X_R)\over {\lambda^4}}\big]^2\,,
\end{eqnarray}
and thus only the additional contributions to $X_{L,R}$ need to be calculated 
within a given New Physics model to obtain constraints. 

Let us quickly survey a few New Physics models which may modify the branching 
fractions for these decays. ($i$) One simple scenario is that of a fourth 
generation~{\cite {fth}} $t'$ which 
gives a contribution $\Delta X_L={\lambda_{t'}\over {\lambda_t}}X_{t'}$ with 
$1.5\leq X_{t'}\leq 5$, depending on the mass of $t'$. This implies that 
${\lambda_{t'}\over {\lambda_t}}\geq \lambda^2-\lambda$ to make any reasonable 
contribution and, hence, is sensitively dependent on what is assumed for the 
generalized CKM matrix structure. ($ii$) A second example is the 
set of Two-Higgs Doublet Models~{\cite {bhp}} with natural flavor 
conservation. Here 
it is well-known that the new contributions to $X_L$ beyond the SM are small, 
of order a few per cent, 
for all charged Higgs masses and values of $\tan \beta$ 
when other experimental constraints are imposed on the model. 
($iii$) Various anomalous $WWZ$ couplings consistent with LEP and 
Tevatron direct bound measurements have been 
shown to make only a small~{\cite {bird}} contribution to $X_L$. 
($iv$) Any leptoquark~{\cite {lepto}} mediating these decay modes 
through tree-level exchange has been shown to be quite heavy, \ie, 
$M_{LQ}>(25-50)\sqrt {\lambda \lambda'}$ TeV, depending on the detailed 
nature of the leptoquark. ($v$) $R$-parity and lepton-number violating 
$\tilde d_{L,R}$ exchanges in the $t$-channel have been shown~{\cite {candp}} 
to contribute to both $X_{L,R}$ and allow for final states with mixed neutrino 
flavors so that the $X_{L,R}$'s would in this case pick up a pair of generation 
labels! For $\tilde d$ masses of order 100 GeV, the E-787 result forces the 
products of the relevant Yukawa couplings in this case to be 
$\leq 10^{-(4-5)}$.

As a last example we consider the case of $R$-parity conserving 
SUSY~{\cite {susy}}. There are many potential contributions to the decay 
amplitude to consider arising from both boxes and $Z$-penguins containing 
SUSY partners in addition 
to those arising from the usual charged Higgs/top exchange. These include the 
intermediate states of ($i$) $\tilde d_i \oplus \tilde g$, 
($ii$) $\tilde d_i \oplus \tilde \chi^0_j$ and 
($iii$) $\tilde u_i \oplus \tilde \chi^+_j$. In addition, box diagrams will 
now also involve $\tilde \ell$ and $\tilde \nu$ exchanges but $\tilde g$ boxes 
are absent due to color conservation.  As is well-known~{\cite {susy}}, unlike 
in the case of $|\Delta S|=2$ transitions, the chargino-squark contributions 
are expected to be dominant. Furthermore, penguins are generally found to 
dominate over contributions from boxes. 

Due to the complexity of the general SUSY parameter space certain 
approximations are employed in evaluating the various contributions; these 
center on the structure of the $6\times 6$ squark mass matrices. 
Conventionally, these matrices are written down in the so-called super-CKM 
basis in which the flavor structure of the quark-squark-gaugino vertices is 
the same as in quark-quark-gauge boson vertices. 
As discussed~{\cite {susy}} by Buras \etal, the calculation simplifies 
by a change to a different basis where $d^i_L\tilde d^j_L \chi^0_n$ 
and $d^i_L\tilde u^j_L \chi^\pm_n$ are flavor diagonal and the 
$d^i_L\tilde u^j_R \chi^+_n$ flavor structure is described by the CKM matrix. 
This basis is arrived at from the super-CKM basis by rotating the up-type 
left-handed squark fields as $\tilde u_L \to V^\dagger \tilde u_L$, with 
$V$ bing the CKM matrix. 
In this case the flavor change in the left-handed sector occurs through the 
non-diagonality of the sfermion propagators. If the off-diagonal 
terms($\Delta$) in the sfermion mass matrices are small in comparison to the 
average sfermion mass, $\tilde m$, it is possible to expand these fermion 
propagators in powers of $\delta=\Delta/\tilde m^2$. The remaining issue is 
then to know how many powers of $\delta$ to keep in the evaluation of the 
various diagrams. Until recently, it has been believed (see Nir and Worah and 
Buras \etal ~{\cite {susy}}) that a single mass-insertion 
approximation~{\cite {hall}} is sufficient for obtaining reasonable estimates 
for the SUSY contributions to the $K^+\to \pi^+\nu \bar \nu$ and 
$K_L\to \pi^0 \nu \bar \nu$ decay amplitudes. In that approximation, it has 
been shown that departures by factors of $2\sim 3$ from the SM decay rate 
expectations are possible once other constraints on the SUSY parameter space 
are taken into account. 

More recently, Colangelo and Isidori~{\cite {susy}} 
have argued that this single mass insertion approximation is not sufficient 
to account for all potentially large SUSY effects and that one must include 
at least two mass insertions. (In their approach one also perturbatively 
expands the matrices responsible for diagonalizing the chargino mass matrix 
to keep track of powers of $SU(2)_L$ breaking terms.) They find that the 
dominant SUSY contribution to $X_L$ can be quite large and is given by 
$\simeq {1\over {96}} {\tilde \lambda_t \over {\lambda_t}}$ where 
$\tilde \lambda_t=(\delta^U_{LR})^*_{ts}(\delta^U_{LR})_{td}=
|\tilde \lambda_t|e^{i\tilde \theta_t}$ and the $\delta$'s are given by the 
$LR$ elements of the $u$-squark mass matrix in the new basis: 
$(\delta^U_{LR})^*_{ab}=[M(LR)^2_U]_{ab}/m^2_{\tilde {u_L}}$. These authors 
then examine all of the other low-energy constraints on the parameter 
$\tilde \lambda_t$ from $K$, $D$, and $B$ physics while simultaneously 
evaluating its 
maximum contribution to $X_L$. One finds that the branching fraction for 
$K^+\to \pi^+\nu \bar \nu$ can be increased by as much as 
an order of magnitude while 
that for $K_L\to \pi^0 \nu \bar \nu$ depends strongly on the new physics phase 
$\tilde \theta_t$. For large values of $\tilde \theta_t$ near $90^o$ the 
enhancement in the branching fraction for this mode may be almost as large 
as two orders of magnitude, which is a very exciting prospect. 

It is clearly quite important to verify and improve upon the E-787 result for 
the $K^+\to \pi^+\nu \bar \nu$ branching fraction and to try to observe the 
decay $K_L\to \pi^0 \nu \bar \nu$ as soon as possible.

\section{Summary}

Rare $K$ decays involving leptons do indeed provide a complementary window to 
direct collider searches as probes of New Physics beyond the Standard Model.

\section*{Acknowledgments}

The author would like to thank J.L. Hewett for discussions related to strange 
physics. He would also like to thank Tony Thomas, Xiao-Gang He and friends for 
their hospitality and for providing such a stimulating atmosphere for this 
Workshop.

%
%%%%%%%%%%%%%%%%%%--- References
%%%%%%%%%%%%%%%%%%%%%%%%%%%%%%%%%%%%%%%%%%%%%%%%%%%%%%%
\def\MPL #1 #2 #3 {Mod. Phys. Lett. {\bf#1},\ #2 (#3)}
\def\NPB #1 #2 #3 {Nucl. Phys. {\bf#1},\ #2 (#3)}
\def\PLB #1 #2 #3 {Phys. Lett. {\bf#1},\ #2 (#3)}
\def\PR #1 #2 #3 {Phys. Rep. {\bf#1},\ #2 (#3)}
\def\PRD #1 #2 #3 {Phys. Rev. {\bf#1},\ #2 (#3)}
\def\PRL #1 #2 #3 {Phys. Rev. Lett. {\bf#1},\ #2 (#3)}
\def\RMP #1 #2 #3 {Rev. Mod. Phys. {\bf#1},\ #2 (#3)}
\def\ZPC #1 #2 #3 {Z. Phys. {\bf#1},\ #2 (#3)}
\def\EPJ #1 #2 #3 {Eur. Phys. J. {\bf#1},\ #2 (#3)}
\def\IJMP #1 #2 #3 {Int. J. Mod. Phys. {\bf#1},\ #2 (#3)}

\end{document}